# Topological insulator-based Dirac hyperbolic metamaterial with large mode indices

Zhengtianye Wang[1,2], Saadia Nasir[3], Yongchen Liu[1], Sivakumar Vishnuvardhan Mambakkam[1], Mingyu Yu[1], Stephanie Law[1,3,*]

1. Department of Material Science and Engineering, University of Delaware, Newark, DE, 19716, U.S.A.
2. Two-dimensional Crystal Consortium, Pennsylvania State University, State College, PA, 16802, U.S.A.
3. Department of Physics and Astronomy, University of Delaware, Newark, DE, 19716, U.S.A.

Corresponding author: slaw@udel.edu

**Abstract**

Hyperbolic metamaterials (HMMs) are engineered materials with a hyperbolic isofrequency surface, enabling a range of novel phenomena and applications including negative refraction, enhanced sensing, and subdiffraction imaging, focusing, and waveguiding. Existing HMMs primarily work in the visible and infrared spectral range due to the inherent properties of their constituent materials. Here we demonstrate a THz-range Dirac HMM using topological insulators (TIs) as the building blocks. We find that the structure houses up to three high-wavevector volume plasmon polariton (VPP) modes, consistent with transfer matrix modeling. The VPPs have mode indices ranging from 126 to 531, 10-100x larger than observed for VPP modes in traditional media while maintaining comparable quality factors. We attribute these properties to the two-dimensional Dirac nature of the electrons occupying the topological insulator surface states. Because these are van der Waals materials, these structures can be grown at a wafer-scale on a variety of substrates, allowing them to be integrated with existing THz structures and enabling next-generation THz optical devices.

**Main text**

Hyperbolicity in solid material optics is of significant interest as it allows one to beat the diffraction limit and paves the way for subdiffraction imaging and waveguiding, negative refraction, enhanced sensing, and other unusual optical behavior. When light propagates inside hyperbolic media, the iso-frequency surface is a hyperboloid, as shown in Fig. 1(a). In these materials, the wavevector $k$ of light that can propagate can be extremely large. Hyperbolic materials exist in nature[1–8]; the origin of the optical anisotropy can include phonon resonances, anisotropic effective masses, and so on. However, natural hyperbolic materials are difficult to tune as it is hard to change their inherent physical properties. Fortunately, there also exist artificial hyperbolic metamaterials (HMMs) comprising subwavelength metal and dielectric components such as metal nanowires embedded in a dielectric matrix or a superlattice of alternating metal and dielectric layers[9,10]. HMMs overcome the tunability issue since the optical properties of the entire system can be modified by tailoring the components, including the type of the metal or dielectric, the optical component geometry (layer thickness or nanowire diameters), the metal-dielectric ratio, and so on. These materials act as effective media with anisotropic permittivities in which either the parallel or perpendicular component of the permittivity is negative while the other is positive, leading to the aforementioned hyperbolic isofrequency surface. The hyperbolicity in layered HMMs arises from the coupling of surface plasmon polaritons at every metal/dielectric interface. These coupled modes go by a variety of names; we will refer to them as volume plasmon polariton (VPP) modes. These VPP modes are high-$k$ bulk cavity modes that comprise the hyperbolic isofrequency surface. They can slow light and increase light-matter interactions, and they are the foundation of subdiffraction imaging, focusing, and waveguiding[11].

HMMs have been explored in the visible and near-infrared ranges using materials like gold, $Al_2O_3$, Si, and $TiO_2$, and in the mid-infrared with heavily doped III-V semiconductors[12–21]. However, despite the many applications, HMMs are rarely investigated in the far-infrared or terahertz (THz) ranges due to a lack of suitable materials. Fortunately, Dirac materials with a high density of states in two dimensions (2D) and low loss like graphene and topological insulators (TIs) are promising as THz HMMs[22]. These HMMs function through the coupling of 2D plasmon polaritons to create the VPP modes that form the hyperbolic isofrequency surface. Dirac HMMs based on graphene have been theoretically investigated for a variety of applications, including as biosensors, switchable reflection modulators, terahertz emitters with large Purcell effect, perfect absorbers, and super resolution lenses[23–31]. However, fabricating such structures at a wafer scale is challenging[32], since the graphene typically must be transferred one layer at a time which is not truly scalable[33]. The direct growth of a graphene multilayer structure is even more challenging due to the difficulty of growing graphene at a wafer scale on a suitable dielectric material. On the other hand, HMMs based on topological insulators (TI) can be readily grown at a wafer scale with molecular beam epitaxy (MBE). A TI is a topological state of matter with 2D massless conducting surface states at the boundaries between the TI and any topologically-trivial material. The plasmons in a TI exist at every interface, comprise Dirac electrons, are 2D and massless, and in general, are in the mid-infrared and THz range[34–48]. In a superlattice comprising alternating TI and trivial band insulator layers, topological surface states hosting Dirac plasmons emerge at each interface. If the TI and band insulator layers are thin enough, these plasmons can couple to each other, forming VPP modes and resulting in a Dirac HMM[49,50].

In this paper, we report numerical simulations and experimental evidence for VPP modes in a superlattice made of the TI $Bi_2Se_3$ and the structurally-compatible trivial band insulator $(Bi_{0.5}In_{0.5})_2Se_3$ (BIS)[51,52]. We grow the multilayer film via MBE and fabricate micro-ribbon structures to couple the light and excite the VPP modes within the system. The dispersion of the VPP modes can be mapped out by analyzing the peaks in the extinction spectra as a function of ribbon width. The experimental dispersion of the VPP modes in our samples matches that predicted by semi-classical transfer matrix method modeling. Finally, we find that the VPP modes show mode indices 10-100x greater than those found in traditional HMMs with comparable quality factors. This is the first experimental demonstration of VPP modes in any Dirac metamaterial and showcases the richness of this TI-based system. The ability to grow this structure at the wafer scale makes it feasible for future large-scale THz-based applications. Because these structures are built from van der Waals materials, they can be grown on a variety of substrates and integrated with existing THz emitters, detectors, or other optoelectronic structures leading to new, efficient, on-chip THz devices[53,54].

**Results and Discussion**:

The multilayer structure used in the study is shown in Fig. 1(b). We use BIS as the buffer layer, capping layer, and spacer layers. Ten topologically-protected surface states are expected to exist within the five-periods of 50 nm $Bi_2Se_3$ and 50 nm BIS. With a thick enough spacer and high enough band offset, these surface states are quantum-mechanically uncoupled but electromagnetically coupled. We will refer to this structure as '5L-50nm'. We first investigate the polariton dispersion using transfer matrix modeling. The color plot in Fig. 1(c) presents the modeling of the 5L-50nm structure, assuming that each topological surface state has a chemical potential $\mu$=0.28 eV and a carrier lifetime $\tau$=1 ps. Below 10 THz, we see one bright branch with a dimmer and almost linearly dispersed branch below it. The brightest branch is the first VPP mode (VPP1) and the second branch is the second VPP mode (VPP2), as labeled in Fig. 1(c). The VPP3 mode appears as another dim linear branch at an even lower frequency which is difficult to see. The bright horizontal dispersionless mode near 5THz corresponds to the epsilon near zero (ENZ) mode of BIS[38] which we will elaborate on in the following. The VPP modes in this region have a group velocity $d\omega/dk >0$, so the HMM is classified as type II. In the 10-21 THz region, we see many branches with group velocity $d\omega/dk <0$, corresponding to a type I HMM.

We have used several assumptions in the modeling. First, the BIS permittivity is obtained from fitting transmission spectra of thick films on sapphire in the 1.5-8 THz range where the sapphire is transparent; we have extrapolated its permittivity to the 8-21 THz range for the model[38]. Second, we used a constant permittivity for silicon across this entire frequency window. The results of the TMM modeling of the 5L-50 nm structure on silicon are close to the results of modeling on sapphire shown in Fig. 4. This indicates that the hyperbolicity of the structure is due to the $Bi_2Se_3$ and BIS layered structure rather than any interaction with the substrate.

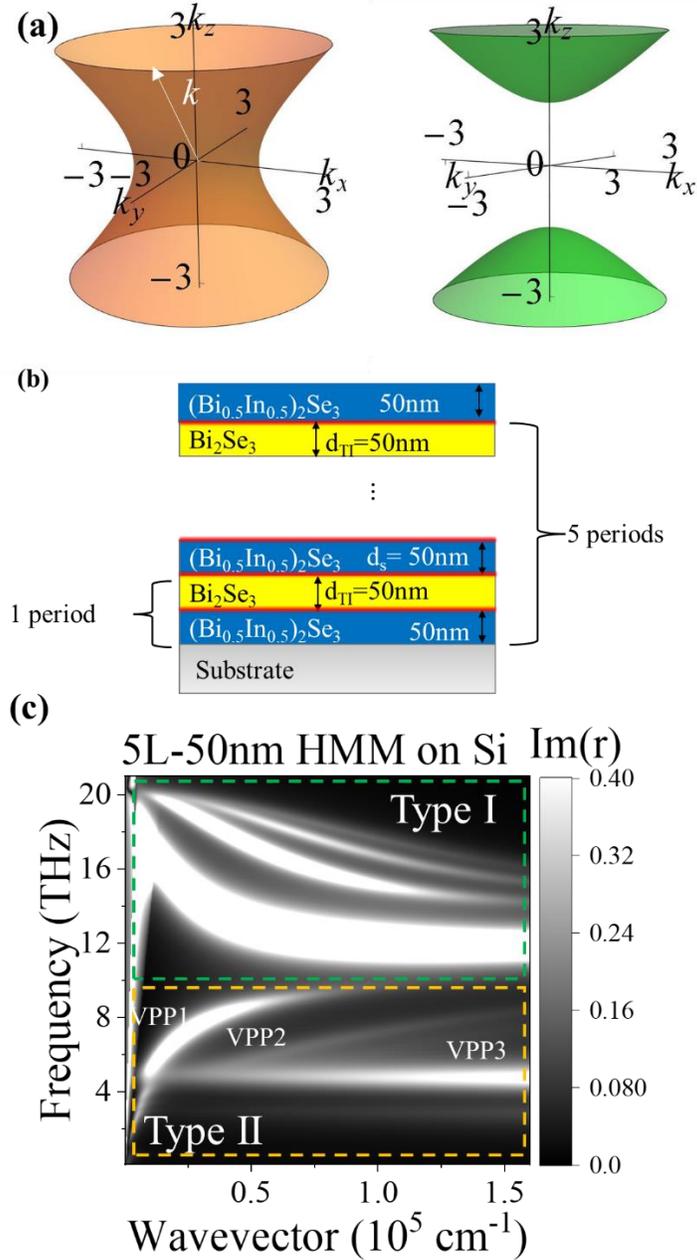

Figure 1 (a) Schematic iso-frequency surfaces of Type II (left) and Type I (right) hyperbolic metamaterials (HMMs). (b) Schematic of the 5L-50nm multilayer HMM structure comprising five pairs of $Bi_2Se_3$ and BIS with an extra BIS capping layer. (c) Transfer matrix modeling of the 5L-50nm $Bi_2Se_3$-BIS HMM. The color plot shows the imaginary part of the Fresnel reflection coefficient calculated for the structure. In the range of 1-9 THz, the structure behaves as a type II HMM, with first volume plasmon polariton (VPP1), second volume plasmon polariton (VPP2), and third volume plasmon polariton (VPP3) existing in the range of 5-9 THz. The non-dispersive mode near 5 THz is the epsilon-near-zero (ENZ) mode of BIS. In the range of 9-20 THz, the structure acts as a type I HMM.

We will investigate the VPP modes in the Type II region where the high-resistivity silicon substrate is sufficiently transparent. After the films are grown using MBE, they are patterned into periodic micro-ribbon arrays using standard photolithography and etching techniques to support the excitation of the VPP modes with large wavevector (for details, see Methods). The polariton wavevectors are given by $k=\pi/a$, where $a$ is the width of the ribbon. Extinction spectra are measured using Fourier transform infrared spectroscopy; only TM-polarized light will excite the VPP modes. By varying the ribbon widths from 3 µm to 0.25 µm, we can map the polariton dispersion. The extinction curves of these samples are shown in Fig. 2.

We first look at the extinction of the sample with $a=3$ µm, shown as the dark blue curve. The most intense peak is centered around 4.7 THz with two kinks on each side at about 3.8 THz and 6.1 THz, respectively. The peak is attributed to the plasmon absorption while the kinks are the result of the Fano interaction between the plasmon and the β phonon or the BIS epsilon-near-zero (ENZ) mode. A Fano interaction happens when a dispersive continuum state (for example, a plasmon) strongly interacts with non-dispersive states with a narrow linewidth (for example, a phonon). This interaction results in an asymmetric extinction curve. In these samples, there is also a Fano interaction between the plasmon and the α phonon, which typically appears near 2 THz and is clearly visible in the TE-polarized extinction spectra (see the Supplementary Information). The strong interaction has caused the peak associated with the α phonon to shift to frequencies below 2 THz where the FTIR has poor sensitivity. The missing α phonon peak at 2 THz is often regarded as an indicator that a plasmon has been excited in this type of sample, though it is not conclusive evidence.

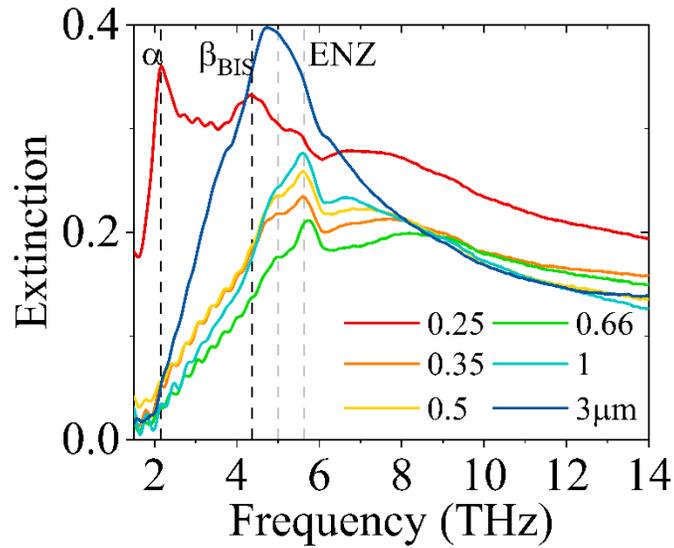

Figure 2 Transverse magnetic (TM) extinction spectra for the 5L-50nm HMM on silicon with ribbon widths from 0.25 μm to 3 μm. The black dashed lines mark the α and β phonons. The grey dashed lines mark the BIS epsilon-near-zero (ENZ) modes.

In samples with ribbon width $a$=1, 0.66, 0.5, 0.35 μm (cyan, green, yellow, and orange curves), a prominent narrow peak at around 5.5 THz accompanied by a kink around 5.0 THz are attributed to the BIS ENZ modes (indicated by gray dashed lines). The permittivity of BIS crosses zero once at the first ENZ point ($\omega_{ENZ1}$≈5.2 THz) and again at the second ENZ point ($\omega_{ENZ2}$≈5.5 THz). These ENZ modes are observable when the wavevector in the sample is relatively large. For the sample with ribbon width $a$=3 μm, it is difficult to resolve the ENZ modes since the plasmon is brighter and overlaps them in frequency. The same ENZ modes are also observed in the 5L-50 nm HMM fabricated on sapphire (see Fig. 4). For these samples, the plasmon polariton mode has shifted to higher frequencies and is visible as a broad peak. When the ribbon width decreases to $a$=0.25 μm (red curve), the plasmon polariton shifts to even higher frequency, reducing the coupling with the α phonon and allowing the phonon peak at 2 THz to reemerge. In addition, a phonon peak near 4.3 THz appears which can be attributed to the BIS β phonon.

In general, we see that the plasmon polariton peak for these samples first blue shifts (1 µm to 0.66 µm), then red shifts (0.66 µm to 0.5 µm), then blue shifts again (0.5 to 0.35 µm). This is inconsistent with the expected dispersion from a single polariton mode and indicates the presence of multiple polariton modes. To extract the mode frequencies from the extinction curves, we employ the analytical Fano model described in the Methods. We use a five-oscillator model to fit the data. One oscillator always describes the α phonon. Based on the features of the curve, two oscillators are used either to describe both BIS ENZ modes (for $a$=1, 0.66, 0.5, 0.35 µm), to describe one $Bi_2Se_3$ β phonon and one ENZ mode (for $a$=3 µm), or to describe one BIS β phonon and one ENZ mode (for $a$=0.25 µm). The remaining two oscillators describe plasmon polaritons with frequencies varying from 4 to 10 THz. As shown in Fig. 3(a), the fitting curves and experimental data match well. In the Supplementary Information, we demonstrate that five oscillators are needed and rule out the possibility of single plasmon polariton mode in these samples.

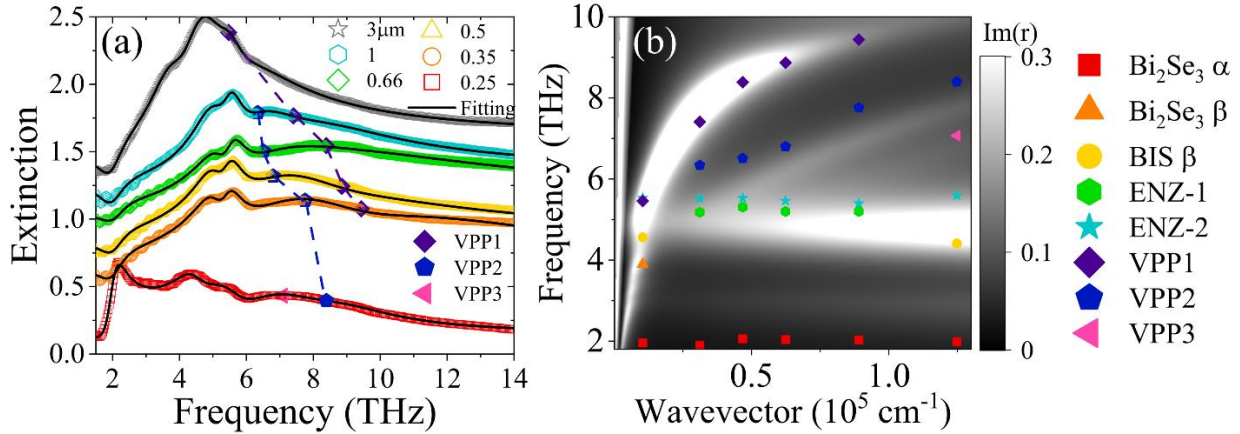

Figure 3 (a) Fano fitting curves for the TM extinction spectra of the 5L-50nm HMM grown on silicon. The data are multiplied by 3 and offset for better visualization. The VPP1 and VPP2 mode frequencies are marked with purple diamonds and blue pentagonsm respectively. VPP3 is marked with a pink triangle. The dashed lines are a guide to the eye. (b) Transfer matrix modeling of the 5L-50nm $Bi_2Se_3$-BIS HMM shown in grayscale. Data points are the corresponding mode frequencies extracted from the Fano fitting.

From the fitting, we can extract the plasmon polariton frequencies. These are shown as solid purple diamonds, blue pentagons, or pink triangles in Fig. 3(a); the dashed lines are a guide to the eye. In Fig. 3(b), we show the transfer matrix modelling results in grayscale. On top of this plot, we overlay the frequencies of all five oscillators used in the Fano modeling. In every sample, we see an oscillator matching the α phonon indicated by a red square. Its frequency varies in a narrow range from 1.90 to 2.06 THz which matches the clear α phonon peak present in the TE extinction spectra shown in the Supplementary Information. In most samples, we see peaks matching the VPP1 (purple diamond) and VPP2 modes (blue pentagon) as well as the ENZ-1 and ENZ-2 modes (green hexagon and cyan star, respectively). Finally, we occasionally see the β phonons associated with the $Bi_2Se_3$ and BIS layers (orange triangle and yellow circle, respectively) as well as the VPP3 mode (pink triangle). We see a good agreement between the many VPP1 and VPP2 data points and the predicted dispersion curves, clearly indicating that this structure is acting like a Dirac HMM in the THz. Although we only observe one VPP3 data point, it too matches well with the predicted VPP3 position. We attribute our inability to observe additional VPP3 data points to the relatively weak intensity of this mode.

To support our claim that the superlattice is acting like an HMM, we can look at data for a superlattice grown on a sapphire substrate. 5L-50 nm multilayer films are directly deposited on a 0.5 mm thick, 1 cm×1 cm sapphire substrate at 300 °C and patterned into ribbon structures with widths $a$=3, 1, 0.66, 0.5, 0.35 and 0.25 µm. The extinction spectra for these samples are shown in Fig. 4(a) where the open, colored symbols are the experimental data points, and the black lines are the Fano fitting. In the Fano fitting, we used the minimum number of oscillators to obtain a good fit, and we confine their frequencies to a small range except for the oscillators associated with the plasmon polaritons. The extracted resonances are overlaid on the color plot generated with TMM modeling in Fig. 4(b). For samples with $a$=3, 1, and 0.66 µm, we need only one plasmon polariton oscillator to fit the data. The resulting plasmon frequencies for the $a$=3 and 0.66 µm samples fall on the dispersion curve for VPP1, while the frequency for the $a$=1 µm sample falls on the dispersion curve for VPP1. For samples with narrower ribbon width, we do not clearly see a peak in the extinction spectra, so we only fit the data with the non-dispersive polariton modes. We attribute our inability to clearly observe multiple VPP modes in this sample to the narrow transparency window of sapphire in this frequency range. The transmission through our sapphire substrates begins to decrease around 7 THz which makes it challenging to distinguish multiple points for VPP1 and VPP2. In addition, the smaller transparency of sapphire compared to silicon makes it difficult to pick out weaker modes such as VPP3. Despite these challenges, these data support our argument that this type of superlattice acts as a Dirac HMM in the THz. If we compare the experimental extinction curves shown in Figure 2 and Figure 4, we see a close resemblance in the shape of curves for samples with the same ribbon widths. This consistency matches our expectation that structurally similar samples should show similar extinction spectra and further confirms that we are observing VPP modes in a Dirac HMM.

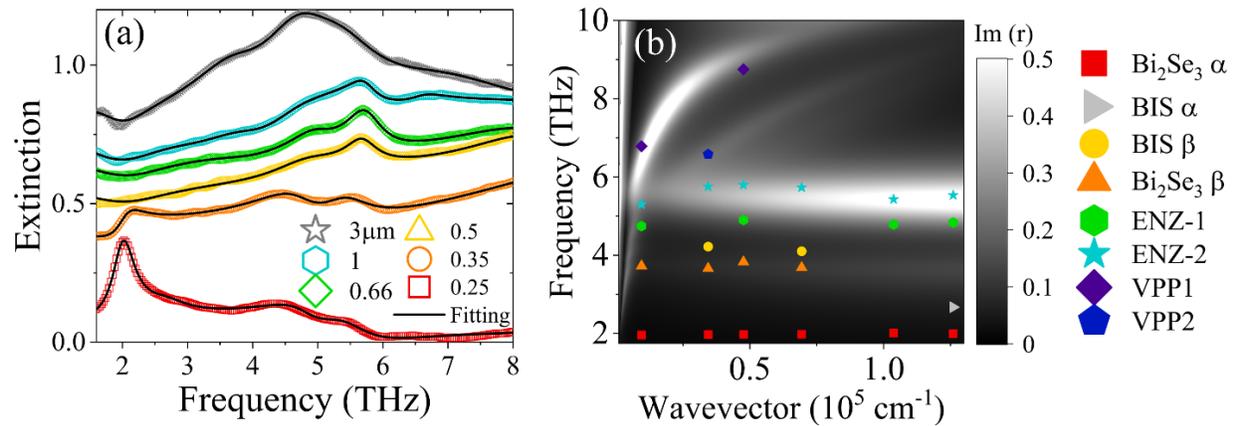

**Figure 4**. (a) Fano fitting curves (black) for the TM extinction spectra (colored symbols) of the 5L-50nm HMM grown on sapphire. The data are offset for better visualization. (b) Transfer matrix modeling of the 5L-50nm $Bi_2Se_3$-BIS HMM shown in grayscale. Data points are the corresponding mode frequencies extracted from the Fano fitting.

In addition to polariton frequencies, the Fano fitting also allows us to obtain the linewidths of these modes. For the VPP modes in the 5L-50nm sample grown on Si, the linewidths vary from 0.8-3.7 THz as shown in Table S1 in the Supplementary Information. This somewhat broad linewidth contributes to the difficulty of distinguishing the two separate VPP modes. The corresponding quality factors ($Q$) of the VPP modes (defined as the mode frequency divided by its linewidth) range from 2.19-6.7, as shown in Table 1. It is likely that with decreasing temperature and/or an improvement in material quality, the carrier lifetime and thus the plasmon lifetime will lengthen, reducing the linewidth of these modes and increasing their quality factor. Nevertheless, these quality factors are comparable to those shown in HMMs made from other materials[11]. Finally, we can calculate the effective index ($n^*$) of the VPP modes, which is a measure of how well they confine light and is critical for applications such as subdiffraction focusing and sensing. We observe mode indices ranging from 126 to 531, 10-100x larger than those observed in other HMMs[55]. These results indicate that this Dirac HMM can confine light up to 100x better than HMMs made from traditional materials while maintaining comparable quality factors. Similarly large mode indices were observed for Dirac plasmon polaritons in TI thin films[36]. The exceptionally large mode indices may be attributable to the two-dimensional nature of the surface plasmon polaritons that comprise the VPP modes. The fact that the material can maintain an acceptable quality factor even with such large mode indices may be due to the reduction in back-scattering for the electrons occupying the spin-polarized TI surface states.

**Table 1**: Mode indices ($n^*$) and quality factors (Q) for the VPP modes of the 5L-50nm HMM on Si

| a (μm) | VPP1 | | VPP2 | | VPP3 | |
|---|---|---|---|---|---|---|
| | $n^*$ | Q | $n^*$ | Q | $n^*$ | Q |
| 3 | 55 | 6.7 | - | - | - | - |
| 1 | 126 | 3.22 | 147 | 4.40 | - | - |
| 0.66 | 167 | 2.25 | 215 | 2.19 | - | - |
| 0.5 | 211 | 3.17 | 276 | 2.52 | - | - |
| 0.35 | 283 | 2.56 | 345 | 2.44 | - | - |
| 0.25 | - | - | 445 | 3.98 | 531 | 3.88 |

**Summary**

In conclusion, we have demonstrated the existence of multiple VPP modes in a five-layer TI/BI superlattice. These modes clearly indicate that this structure is acting like a Dirac HMM in the THz spectral range. We show that this structure can be grown on semiconductor substrates with dangling bonds like silicon as well as oxide substrates like sapphire with similar quality. The VPP modes in these structure have mode indices 10-100x larger than those found in HMMs made from traditional materials with comparable quality factors. In the future, the position and dispersion of the VPP modes could be tuned by adjusting the material structural parameters or through gating the TI layers, unlike natural hyperbolic materials which are difficult to tune. These extremely large mode indices combined with the fact that these structures can be grown on a variety of substrates by van der Waals epitaxy opens the door to the creation of new types of integrated, on-chip optical and optoelectronic devices in the THz spectral range including on-chip sensors or hyperlenses to enable subdiffraction imaging.

**Acknowledgements:** Z.W., S.N., and S. L. acknowledge funding from U.S. Department of Energy, Office of Science, Office of Basic Energy Sciences, under Award No. DE-SC0017801. Y.L. acknowledges funding from UD-CHARM, a National Science Foundation MRSEC, under Award No. DMR-2011824. S.V.M. acknowledges funding from the National Science Foundation, Division of Materials Research, under Award No. 1838504. M.Y. acknowledges support from the II-VI Foundation Block Grant Program. The authors acknowledge the use of the Materials Growth Facility (MGF) at the University of Delaware, which is partially supported by the National Science Foundation Major Research Instrumentation under Grant No. 1828141 and UD-CHARM, a National Science Foundation MRSEC, under Award No. DMR-2011824.

**Author contributions:** Z.W. and S.L. conceived the idea, designed the experiment, and wrote and revised the manuscript. Z.W. carried out the material growth, nanofabrication, FTIR measurements, data analysis, and transfer matrix modeling. S.N., Y.L., and S.V.M. assisted with nanofabrication and structural characterization. M.Y. assisted with substrate preparation and atomic force microscopy confirmation of ribbon widths. S.L. supervised the project. All authors discussed the results and commented on the manuscript.

**Competing interests:** The authors declare no competing interests.

**Methods**

**Epitaxial growth:** The $Bi_2Se_3$/BIS multilayer structures were grown on 1 cm×1 cm silicon (111) substrates via MBE using a Veeco GENxplor system. The substrate (MTI Corp.) is first cleaned via the RCA cleaning process (for details, see Supporting Information) and immediately baked at 200 °C in the load lock for 12 hours under high vacuum. After being transferred to the growth chamber (base pressure<$10^{-9}$ Torr), a 1×1 surface reconstruction is observed in the reflection high energy electron diffraction (RHEED) pattern with blurry diffraction streaks. The substrate is further cleaned by flashing it to 900 °C with ramp rate of 70 °C/min to obtain a clearer RHEED pattern, then cooled to 100 °C for deposition. Dual zone Knudson effusion cells are used for bismuth (UMC Corp., 6N purity grade), and indium (UMC Corp., 7N purity grade) sources. For the selenium cracking cell (UMC Corp., 6N purity grade), the base temperature is kept between 280-290°C while the cracker zone kept at 900°C. The bismuth bulk base temperature ranges from 470 °C-490 °C, depending on the growth rate needed. The cell tip temperature is kept 25 °C higher than the bulk temperature to prevent condensation near the tip. Similarly, the indium bulk temperature ranges from 670 °C-690 °C, and the tip temperature is kept 100 °C higher. The substrate is held by a rotating manipulator and temperatures are measured by a non-contact thermocouple. The $Bi_2Se_3$ layers are grown at a substrate temperature of 325 °C and the BIS layers are grown at a substrate temperature of 300 °C. A selenium to bismuth flux ratio (defined by beam equivalent pressure as measured by a thoria coated iridium filament) or selenium to bismuth plus indium flux ratio is kept at 70 to maintain stoichiometry of the film. Details of the growth are included in the Supporting Information. After the final layer is grown, the sample is cooled to 200 °C in a selenium flux before being transferred out of the growth chamber. The films were stored in a vacuum pack to await measurement and fabrication. The growth process on sapphire was identical, except the substrate was heated to 650°C in the chamber before growth to desorb any contamination.

**Ribbon Fabrication.** To fabricate the micro-ribbon structures, we used electron-beam lithography (EBL) and argon ion milling. We used AR-P 6200.09 resist (Allresist, DE) with a thickness of 200nm. The narrowest ribbon width used in this project was 250nm; the filling ratio was always 50%. After the e-beam exposure (base dose: 200 µC/cm$^2$) and development (developer: AR600-546), the samples were etched with argon ion milling. The remaining resist was removed by soaking the etched samples in N-Methyl-2-Pyrrolidone (NMP) in a hot water bath at 80°C for an hour. The samples were then rinsed with acetone and isopropanol and dried with nitrogen gas.

**FTIR measurement.** The extinction spectra of these samples were measured the day after nanofabrication using a Bruker Vertex 70v FTIR. The extinction of the sample is calculated as E=1-T/T$_0$, where T is the transmission of the sample, and T$_0$ is the transmission of the substrate. Both transverse magnetic (TM) spectra and transverse electric (TE) spectra were measured. For TM measurements, the electric field is perpendicular to the ribbons while for TE measurements, the electric field is parallel to the ribbons. The spectra are taken with 1 cm$^{-1}$ (0.03 THz) resolution and a 1.6 kHz scan rate. The TM spectra are based on an average of 10000 scans while the TE spectra are based on an average of 5000 scans.

**Fano Fitting:** To obtain the plasmon modes frequencies and linewidths, we use the analytical Fano resonance model to describe the interactions between non-dispersive phonons, BIS epsilon-near-zero (ENZ) modes and the VPP modes using the following equations:

$$\text{Extinction}(\omega) = 1 - T(\omega) = |e(\omega)|^2 \tag{1}$$

$$e(\omega) = a_r - \sum_j e_j(\omega) \tag{2}$$

$$e_j(\omega) = \frac{b_j \Gamma_j e^{i\varphi_j}}{\omega - \omega_j + i\Gamma_j} \qquad (3)$$

In these equations, $a_r$ is the background constant, $e(\omega)$ is the overall polarizability of the studied system, and we describe the $j_{th}$ resonator with frequency $\omega_j$, linewidth $\Gamma_j$, phase $\varphi_j$, and amplitude $b_j$. The TM-polarized data from 1.5 to 14 THz are used for the fitting process. The Fano fitting is programmed using the NonlinearModel Fit embedded in Wolfram Mathematica. All of the oscillator parameters $\omega_j$, $\Gamma_j$, $\varphi_j$, $b_j$, and $a_r$ are the fitting parameters, while extinction and frequency $\omega$ are the variables. Some constraints are used for the fitting: for example, the α phonon frequency $\omega_\alpha$ was constrained to lie within 1.8 THz $<\omega_\alpha<$ 2.3 THz, the lower and upper limit defined by the α phonon frequency of pure $Bi_2Se_3$ and BIS respectively. All the fittings in the main text have residual squares R>0.9995.

**Transfer Matrix Modeling:** A transfer matrix method is used to model the optical properties of the heterostructure. The permittivities of $Bi_2Se_3$ and BIS are modeled with a Lorentz model[38]. The silicon substrate is modeled as semi-infinite with a constant permittivity $\varepsilon_{Si}$=11.7. The surface states are modeled as infinitely thin conducting interfaces with conductivity given by: $\sigma=i\mu/4\pi(\omega+i\tau^{-1})$. Here $\mu$=0.28eV is the chemical potential of the surface state electrons which is determined by the measured electron sheet density of $1.2\times10^{13}$ cm$^{-2}$ for a 50 nm $Bi_2Se_3$ film grown on a BIS buffer. We take $\tau$=1 ps. In electrical transport measurement, this carrier lifetime corresponds to a mobility of $\mu_m$=10$^4$ cm$^2$/(V·s).

**References:**
1. Dai, S. *et al.* Graphene on hexagonal boron nitride as a tunable hyperbolic metamaterial. *Nat. Nanotechnol.* **10**, 682–686 (2015).
2. Guo, Y. & Jacob, Z. Broadband super-planckian thermal emission from hyperbolic metamaterials. *Appl. Phys. Lett.* **101**, 131106 (2012).
3. Sun, J., Litchinitser, N. M. & Zhou, J. Indefinite by Nature: From Ultraviolet to Terahertz. *ACS Photonics* **1**, 293–303 (2014).
4. Narimanov, E. E. & Kildishev, A. V. Metamaterials: Naturally hyperbolic. *Nat. Photonics* **9**, 214–216 (2015).

5. Wu, J. S., Basov, D. N. & Fogler, M. M. Topological insulators are tunable waveguides for hyperbolic polaritons. *Phys. Rev. B* **92**, 205430 (2015).
6. Caldwell, J. D. *et al.* Sub-diffractional volume-confined polaritons in the natural hyperbolic material hexagonal boron nitride. *Nat. Commun.* **5**, 1–9 (2014).
7. Korzeb, K., Gajc, M. & Pawlak, D. A. Compendium of natural hyperbolic materials. *Opt. Express* **23**, 25406 (2015).
8. Caldwell, J. D. *et al.* Sub-diffractional volume-confined polaritons in the natural hyperbolic material hexagonal boron nitride. *Nat. Commun.* **5**, 5221 (2014).
9. Ferrari, L., Wu, C., Lepage, D., Zhang, X. & Liu, Z. Hyperbolic metamaterials and their applications. *Prog. Quantum Electron.* **40**, 1–40 (2015).
10. Kidwai, O., Zhukovsky, S. V. & Sipe, J. E. Effective-medium approach to planar multilayer hyperbolic metamaterials: Strengths and limitations. *Phys. Rev. A* **85**, 053842 (2012).
11. Sohr, P., Wei, D., Tomasulo, S., Yakes, M. K. & Law, S. Simultaneous Large Mode Index and High Quality Factor in Infrared Hyperbolic Metamaterials. *ACS Photonics* **5**, 4003–4008 (2018).
12. Sreekanth, K. V., De Luca, A. & Strangi, G. Excitation of volume plasmon polaritons in metal-dielectric metamaterials using 1D and 2D diffraction gratings. *J. Opt.* **16**, 105103 (2014).
13. Wei, D., Harris, C. & Law, S. Volume plasmon polaritons in semiconductor hyperbolic metamaterials. *Opt. Mater. Express* **7**, 2672–2681 (2017).
14. Zhukovsky, S. V., Kidwai, O. & Sipe, J. E. Physical nature of volume plasmon polaritons in hyperbolic metamaterials. *Opt. Express* **21**, 14982–7 (2013).
15. Sreekanth, K. V., De Luca, A. & Strangi, G. Experimental demonstration of surface and bulk plasmon polaritons in hypergratings. *Sci. Rep.* **3**, 3291 (2013).
16. Smalley, J. S. T., Vallini, F., Shahin, S., Kanté, B. & Fainman, Y. Gain-enhanced high-k transmission through metal-semiconductor hyperbolic metamaterials. *Opt. Mater. Express* **5**, 2300 (2015).
17. Sohr, P. & Law, S. Structural parameters of hyperbolic metamaterials controlling high-k mode resonant wavelengths. *J. Opt. Soc. Am. B* **37**, 3784 (2020).
18. Chandrasekar, R. *et al.* Lasing Action with Gold Nanorod Hyperbolic Metamaterials. *ACS Photonics* **4**, 674–680 (2017).
19. Kim, J. *et al.* Improving the radiative decay rate for dye molecules with hyperbolic metamaterials. *Opt. Express* **20**, 8100 (2012).
20. Naik, G. V., Liu, J., Kildishev, A. V., Shalaev, V. M. & Boltasseva, A. Demonstration of Al:ZnO as a plasmonic component for near-infrared metamaterials. *Proc Natl Acad Sci* **109**, 8834–8838 (2012).
21. Urbas, A. M. *et al.* Roadmap on optical metamaterials. *J. Opt.* **18**, 093005 (2016).
22. Atsushi Ishikawa, T. T. Plasmon hybridization in graphene metamaterials. *Appl. Phys. Lett.* 253110 (2013) doi:10.1063/1.4812813.
23. Othman, M. A. K., Guclu, C. & Capolino, F. Graphene-based tunable hyperbolic metamaterials and enhanced near-field absorption. *Opt. Express* **21**, 7614 (2013).
24. Iorsh, I. V., Mukhin, I. S., Shadrivov, I. V., Belov, P. A. & Kivshar, Y. S. Hyperbolic metamaterials based on multilayer graphene structures. *Phys. Rev. B - Condens. Matter Mater. Phys.* **87**, 075416 (2013).
25. Sreekanth, K. V., De Luca, A. & Strangi, G. Negative refraction in graphene-based hyperbolic metamaterials. *Appl. Phys. Lett.* **103**, 023107 (2013).

26. Cynthia, S., Ahmed, R., Islam, S., Ali, K. & Hossain, M. Graphene based hyperbolic metamaterial for tunable mid-infrared biosensing. *RSC Adv.* **11**, 7938–7945 (2021).
27. Pianelli, A. *et al.* Graphene-based hyperbolic metamaterial as a switchable reflection modulator. *Opt. Express* **28**, 6708–6718 (2020).
28. Kozina, O. N., Melnikov, L. A. & Nefedov, I. S. A theory for terahertz lasers based on a graphene hyperbolic metamaterial. *J. Opt.* **22**, 095003 (2020).
29. Xiang, Y. *et al.* Critical coupling with graphene-based hyperbolic metamaterials. *Sci. Rep.* **4**, 5483 (2014).
30. Zhukovsky, S. V., Andryieuski, A., Sipe, J. E. & Lavrinenko, A. V. From surface to volume plasmons in hyperbolic metamaterials: General existence conditions for bulk high-k waves in metal-dielectric and graphene-dielectric multilayers. *Phys. Rev. B* **90**, 155429 (2014).
31. Nefedov, I. S., Valagiannopoulos, C. A. & Melnikov, L. A. Perfect absorption in graphene multilayers. *J. Opt.* **15**, 114003 (2013).
32. Lin, H. *et al.* A 90-nm-thick graphene metamaterial for strong and extremely broadband absorption of unpolarized light. *Nat. Photonics* **13**, 270–276 (2019).
33. Chang, Y., Liu, C., Liu, C., Zhang, S. & Marder, S. R. Realization of mid-infrared graphene hyperbolic metamaterials. *Nat. Commun.* **7**, 10568 (2016).
34. Nasir, S., Wang, Z., Mambakkam, S. V. & Law, S. In-plane plasmon coupling in topological insulator $Bi_2Se_3$ thin films. *Appl. Phys. Lett.* **119**, 201103 (2021).
35. Pietro, P. Di *et al.* Observation of Dirac plasmons in a topological insulator. *Nat. Nanotechnol.* **8**, 556–560 (2013).
36. Ginley, T. P. & Law, S. Coupled Dirac Plasmons in Topological Insulators. *Adv. Opt. Mater.* **6**, 1800113 (2018).
37. In, C. *et al.* Control over Electron-Phonon Interaction by Dirac Plasmon Engineering in the $Bi_2Se_3$ Topological Insulator. *Nano Lett.* **18**, 734–739 (2018).
38. Wang, Z. *et al.* Plasmon coupling in topological insulator multilayers. *Phys. Rev. Mater.* **4**, 115202 (2020).
39. Autore, M. *et al.* Observation of Magnetoplasmons in $Bi_2Se_3$ Topological Insulator. *ACS Photonics* **2**, 1231–1235 (2015).
40. Autore, M. *et al.* Plasmon-Phonon Interactions in Topological Insulator Microrings. *Adv. Opt. Mater.* **3**, 1257–1263 (2015).
41. Autore, M. *et al.* Terahertz plasmonic excitations in $Bi_2Se_3$ topological insulator. *J. Phys. Condens. Matter* **29**, 183002 (2017).
42. Brey, L. Plasmonics in Topological Insulators: Spin − Charge Separation, the Influence of the Inversion Layer, and Phonon − Plasmon Coupling. (2017) doi:10.1021/acsphotonics.7b00524.
43. Di Pietro, P. *et al.* Plasmonic Excitations in $Bi_2Se_3$ Topological Insulator. *J. Phys. Condens. Matter* **29**, 183002 (2017).
44. Qi, J., Liu, H. & Xie, X. C. Surface plasmon polaritons in topological insulators. *Phys. Rev. B* **89**, 155420 (2014).
45. Profumo, R. E. V., Asgari, R., Polini, M. & MacDonald, A. H. Double-layer graphene and topological insulator thin-film plasmons. *Phys. Rev. B* **85**, 085443 (2012).
46. Efimkin, D. K. & Lozovik, Y. E. Collective excitonic and plasmonic excitations on a surface of 3D topological insulator. *J. Phys. Conf. Ser.* **393**, 012016 (2012).
47. Efimkin, D. K., Lozovik, Y. E. & Sokolik, A. A. Spin-plasmons in topological insulator. *J. Magn. Magn. Mater.* **324**, 3610–3612 (2012).

48. Deshko, Y., Krusin-Elbaum, L., Menon, V., Khanikaev, A. & Trevino, J. Surface plasmon polaritons in topological insulator nano-films and superlattices. *Opt. Express* **24**, 7398 (2016).
49. Zhang, R. Z. Optical and thermal radiative properties of topological insulator semiconductor multilayers. *J. Quant. Spectrosc. Radiat. Transf.* **253**, 107133 (2020).
50. Sreekanth, K. V. & Simpson, R. E. Super-collimation and negative refraction in hyperbolic Van der Waals superlattices. *Opt. Commun.* **440**, 150–154 (2019).
51. Koirala, N. *et al.* Record Surface State Mobility and Quantum Hall Effect in Topological Insulator Thin Films via Interface Engineering. *Nano Lett.* **15**, 8245–8249 (2015).
52. Wang, Y., Ginley, T. P. & Law, S. Growth of high-quality $Bi_2Se_3$ topological insulators using $(Bi_{1-x}In_x)_2Se_3$ buffer layers Growth of high-quality Bi2Se3 topological insulators using (Bi1-xInx)2Se3 buffer layers. *J. Vaccum Sci. Technol. B* **36**, 02D101 (2018).
53. To, D. Q. *et al.* Strong coupling between a topological insulator and a III-V heterostructure at terahertz frequency. *Phys. Rev. Mater.* **6**, 035201 (2022).
54. To, D. Q. *et al.* Surface plasmon-phonon-magnon polariton in a topological insulator-antiferromagnetic bilayer structure. *Phys. Rev. Mater.* **6**, 085201 (2022).
55. Mahmoodi, M. *et al.* Existence Conditions of High-k Modes in Finite Hyperbolic Metamaterials. *Laser Photonics Rev.* **13**, 1800253 (2019).

# Supplementary Information for

# Topological insulator-based Dirac hyperbolic metamaterial with large mode indices

Zhengtianye Wang[1,2], Saadia Nasir[3], Yongchen Liu[1], Sivakumar Vishnuvardhan Mambakkam[1], Mingyu Yu[1], Stephanie Law[1,3,*]

4. Department of Material Science and Engineering, University of Delaware, Newark, DE, 19716, U.S.A.
5. Two-dimensional Crystal Consortium, Pennsylvania State University, State College, PA, 16802, U.S.A.
6. Department of Physics and Astronomy, University of Delaware, Newark, DE, 19716, U.S.A.

Corresponding author: slaw@udel.edu

## I. Growth of the $Bi_2Se_3$ and $(Bi_{0.5}In_{0.5})_2Se_3$ superlattice on silicon

**Substrate preparation**:

There is always a layer of silicon oxide on top the silicon wafer, so it is critical to remove the oxide layer before any epitaxial growth. Because the surface oxide layer on our wafer is thin (~2nm), we used an RCA clean. After removing the 1 cm×1 cm Si (111) substrate from the package, we sonicated it in acetone and isopropanol for 10min each to remove possible organic contamination. Then we dipped the wafer in a base-peroxide mixture (volume ratio: $H_2O$:$H_2O_2$ (30% by weight):$NH_4OH$ (29% by weight)=5:1:1) at 80 °C for 10 min to further remove organics and other contamination (metals, etc.) from the surface. This step is known as the first standard clean step (SC-1). The native oxide dissolves at a very low rate and a new silicon oxide layer (~1nm) regenerates at the same time at approximately the same rate. After rinsing the wafer in flowing water for 1min, we dipped the wafer in an acid-peroxide mixture (volume ratio: $H_2O$:$H_2O_2$ (30% by weight):HCl(37% by weight)=6:1:1) at 80 °C for 10 min. This second standard clean step (SC-2) removes the trace of ions introduced in SC-1. Finally, we rinse the wafer in flowing water for about 1 min, followed by drying with nitrogen flow. The wafer is immediately loaded into the molecular beam epitaxy (MBE) system load lock, which is pumped down below $10^{-6}$ Torr in 2 min. The wafer is baked in the load lock at 200 °C for 12 hrs before loading into the main MBE chamber.

After loading the substrate into the growth chamber, the substrate reflection high energy electron diffraction (RHEED) pattern shows the expected 1×1 Si (111) surface reconstruction, as shown in Fig. S1(a). The substrate is the flashed to 900 °C at a 70 °C/min ramp rate to further clean the substrate. After reaching 900 °C, the substrate heater is turned off and the substrate naturally cools to 100 °C. We note that if the substrate were flashed above 950 °C in our chamber, a 7×7 surface reconstruction appears as shown in Fig. S1(b). When the substrate is cooled to around 710 °C, the Si (111) surface return to 1×1 reconstruction. To minimize the history of the substrate surface, we therefore only flash the substrate to 900 °C.

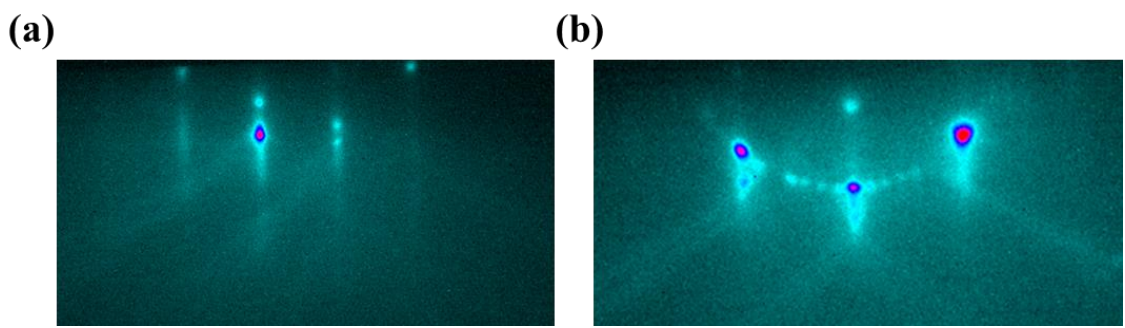

**Figure S1**. RHEED pattern of (a) 1×1 surface reconstruction and (b) 7×7 surface reconstruction on a Si (111) substrate.

### Film Deposition:

Before the $Bi_2Se_3$ deposition, the substrate is exposed to a selenium flux (same beam equivalent pressure as used for the $Bi_2Se_3$ growth step) for 1 min at 100 °C to passivate the surface dangling bonds. Then 3 nm of $Bi_2Se_3$ are deposited at 100 °C, followed by 2 nm of $Bi_2Se_3$ and 5 nm of $In_2Se_3$ at 300 °C and finally 40 nm $(Bi_{0.5}In_{0.5})_2Se_3$ via co-deposition to form a buffer layer. The subsequent $Bi_2Se_3$ and $(Bi_{0.5}In_{0.5})_2Se_3$ layers are deposited at 325 °C and 300 °C, respectively. The manipulator rotates the substrate at 2 °C/min throughout the process to maintain an even substrate temperature and deposition rate across the wafer. A 5 μm×5 μm atomic force microscopy image of the 5L-50nm sample is shown in Fig. S2. The root mean square roughness of the surface is 11.59nm out of a 550 nm film.

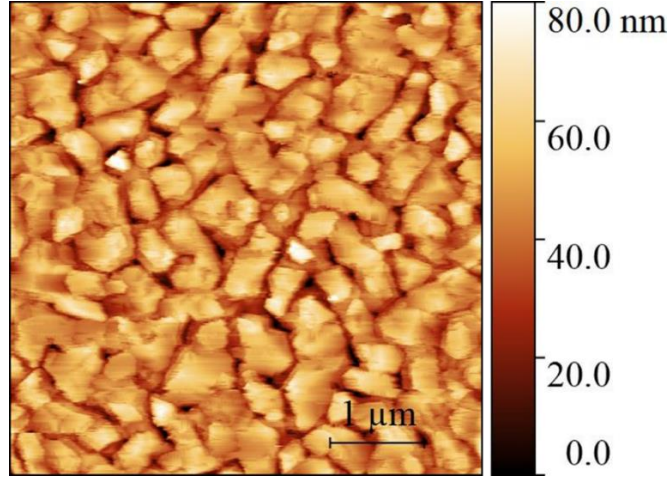

**Figure S2**. Atomic force microscopy image of the 5L-50nm $Bi_2Se_3$ and $(Bi_{0.5}In_{0.5})_2Se_3$ superlattice film on Si (111), size: 5 µm×5 µm. Taken by Veeco Dimension-3100V SPM.

## II.     TE data of 5L-50nm on Si

The TE spectra of the samples are shown in Fig. S3. In the TE set up, the electric field is polarized parallel to the ribbons. The extinction is defined as 1-T/$T_0$, where T is the spectrum of the sample and $T_0$ is the substrate spectrum with the same optical set up. The TE data confirms that the samples remain the same after the nanofabrication process and that the phonons in the system are nearly the same for all six samples.

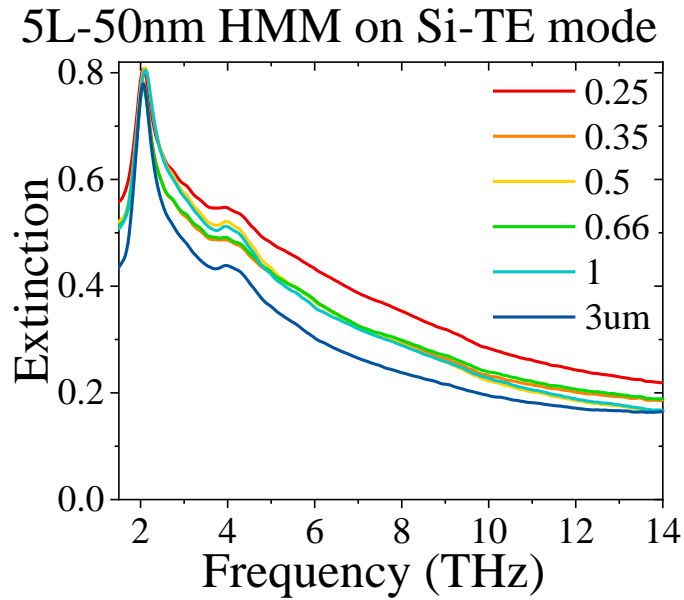

**Figure S3**. TE spectra of the 5L-50nm HMM on Si (111), measured by FTIR with 1.6 kHz scan rate and integrated over 5000 scans.

### III. Simultaneous Four-Oscillator Fitting

Because of the similarities among the extinction curves of samples with $a$=1, 0.66, 0.5, 0.35 µm, we decided to see if we could adopt a simultaneous fitting procedure where we hold the α phonon, ENZ1, and ENZ2 mode constant for all four sets of data and only allow the plasmon to vary. Compared to the independent four oscillators fitting where we have 4 (oscillators) ×4 ($\omega_j$, $\Gamma_j$, $\varphi_j$, $b_j$) =16 fitting parameters for each set of data and 16×4=64 for all 4 sets, the simultaneous fitting has reduced the fitting parameters to 3 (oscillators) ×4 (parameters $\omega_j$, $\Gamma_j$, $\varphi_j$, $b_j$) + 4 (parameters $\omega_j$, $\Gamma_j$, $\varphi_j$, $b_j$) × 4 (data sets) =28.

Fig. S4 shows the simultaneous fitting results with residual square R>0.999. Fig. S4(a-d) are the fitting curves overlaid on top of the experimental data. Good fits are observed for the samples with $a$=0.66 and 0.5 µm. However, for samples with $a$=1 and 0.35 µm, the one free oscillator model cannot capture the curve shape in the range of 6 to 10 THz. Furthermore, we can plot the individual oscillators $|e_j(\omega)|^2$ in Fig. S4(e). The plasmon frequencies obtained via the simultaneous fitting are quite close to each other, in a narrow range from 6.4 to 6.8 THz, which indicates an almost non-dispersive plasmon polariton mode. The fitting result is not consistent with our TMM modeling. The four oscillators simultaneous fitting indicates that an extra free oscillator is needed to fit the TM-polarized extinction spectra.

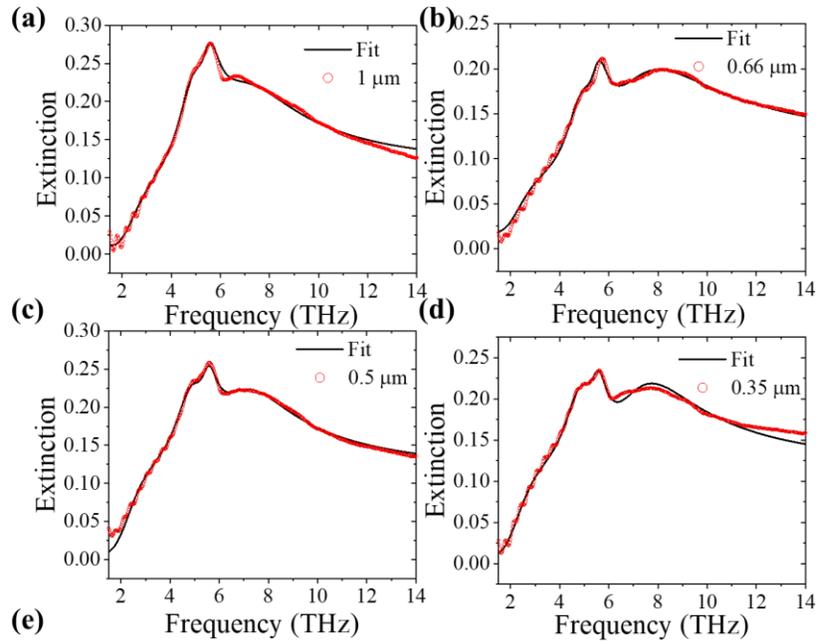
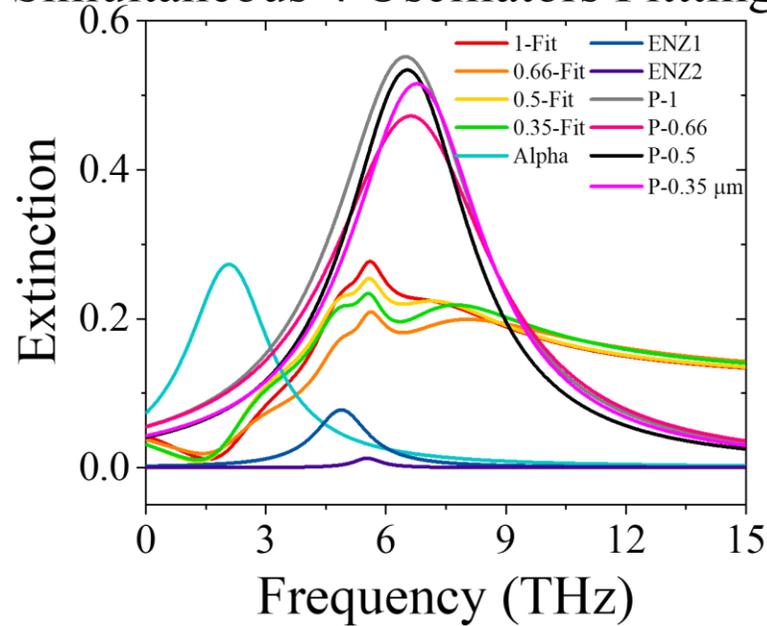

**Figure S4**. (a-d) Experimental data points (open red circles) and fitting curves (black lines) for the simultaneous four oscillator fitting for samples with ribbon widths $a$=1, 0.66, 0.5, 0.35 μm, respectively. (e) Individual oscillators extracted from the Fano fitting process. Red, orange, yellow, and green curves are the fitting curves shown in panels (a-d). Cyan, blue, and violet curves are the α phonon, BIS ENZ1 mode and ENZ2 mode respectively. The gray, black, magenta, and pink curves represent the plasmon polaritons.

## IV. Independent Four-Oscillator Fitting

Because the simultaneous fitting was unable to give satisfying fitting curves for the samples with $a$=1 and 0.35 µm, we returned to the independent fitting but only used four oscillators as opposed to the five-oscillator model in the main text. Fig. S5 shows the best fitting results using this model. For the sample with $a$=0.35 µm, we can get an acceptable fit with a single plasmon frequency at 6.61THz, while for the sample with $a$=1 µm, the shape in the range from 6 to 8 THz is still not captured by the fitting curve (though it is better than the fitting in the simultaneous fitting process shown in Fig. S4(a)). The single plasmon frequency is at 6.17 THz. We did not do the independent fitting for the other two samples as the fitting curve in the simultaneous fitting process already matches the experimental data well. We would expect similar results coming from independent fitting.

Even with the independent four-oscillator fitting process, the plasmon frequencies are within 1THz, which is almost non-dispersive. In addition, the curve shape for the $a$=1 µm sample cannot be captured using either fitting method. This indicates that the single plasmon assumption is not applicable to these samples. Thus, we adopted the five-oscillator model as described in the main text to capture the expected two plasmon modes.

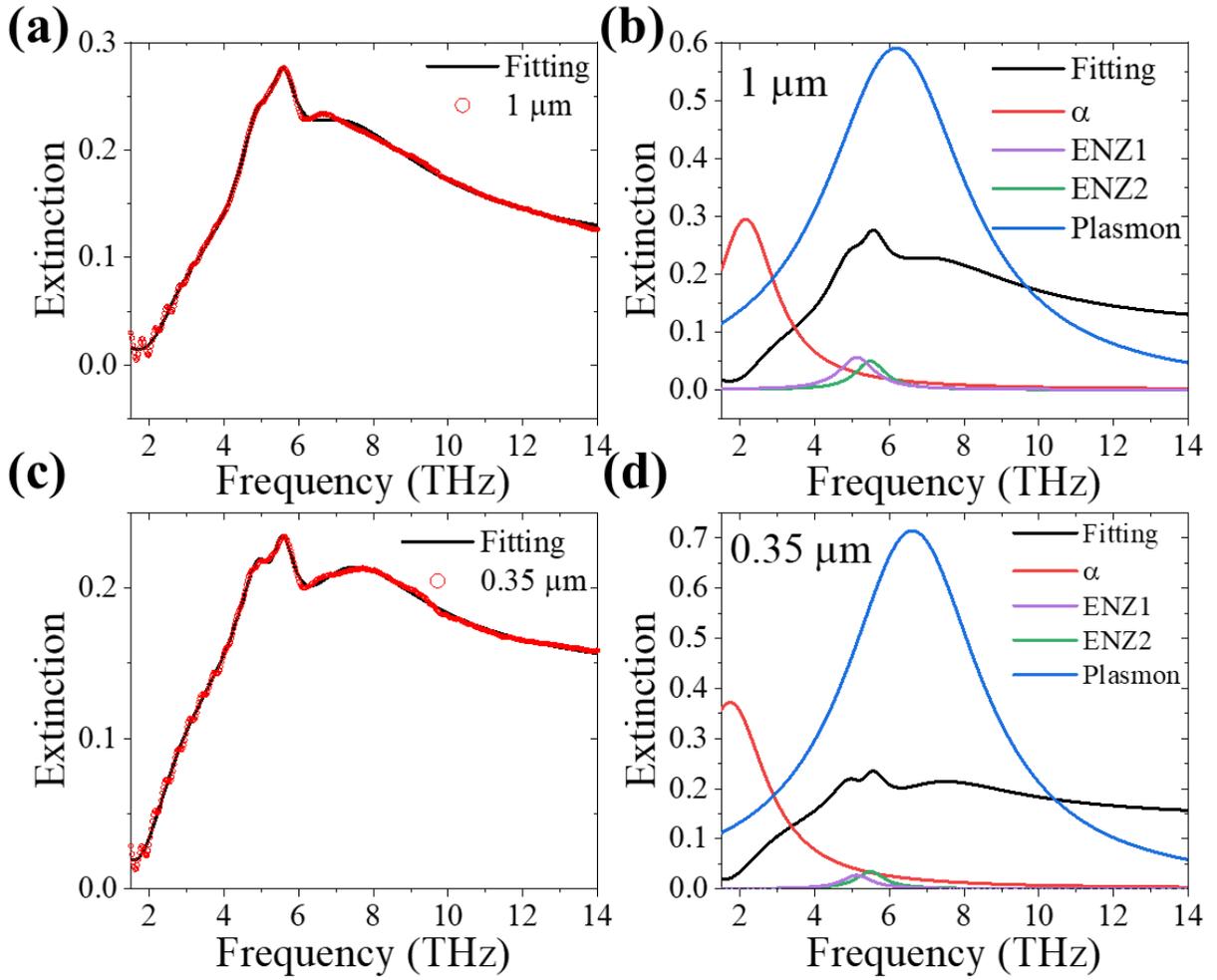

**Figure S5**. Experimental data (red open circles) and independent four-oscillator Fano fitting curve (black solid lines) for samples with $a=1$ (a) and 0.35 µm (c). Individual oscillators extracted from the Fano fitting process for samples with $a=1$ (b) and 0.35 µm (d). Red, violet, green, and blue curves are the α phonon, BIS ENZ1 mode, ENZ2 mode, and plasmon mode, respectively. The black curves are the Fano fitting curves from (a, c).

V.       Fitting Parameter Tables

**Table S1.** Fitting parameters for the five-oscillator model for the 5L-50nm sample on silicon in the main text. $a_r$ is a background parameter, $\omega$, $b$, $\gamma$, and $\varphi$ represent the frequency, strength, linewidth, and phase of the oscillators, respectively. α, BISβ, and BSβ represent the α phonon of the system, the β phonon of BIS, and the β phonon of $Bi_2Se_3$, respectively. e1 and e2 represent

the ENZ1 and ENZ2 modes of BIS, respectively. v1, v2, and v3 represent the VPP1, VPP2, and VPP3 modes of the HMM. Q indicates the quality factor and n$^*$ the mode index of the VPP modes.

| | 3 μm | | 1 μm | 0.66 μm | 0.5 μm | 0.35 μm | | 0.25 μm |
|---|---|---|---|---|---|---|---|---|
| a$_r$ | 0.30917 | a$_r$ | 0.277941 | 0.263531 | 0.271318 | 0.225 | a$_r$ | 0.432525 |
| ω-α | 1.96284 | ω-α | 1.97324 | 2.06398 | 2.03902 | 2.033 | ω-α | 1.9915 |
| b-α | 0.231025 | b-α | 0.401679 | 0.409229 | 0.118049 | 0.473457 | b-α | 0.17302 |
| γ-α | 0.462118 | γ-α | 0.477794 | 0.495348 | 0.471893 | 0.485169 | γ-α | 0.30545 |
| φ-α | 3.57034 | φ-α | 2.96042 | 3.66412 | 3.65271 | 3.72014 | φ-α | 4.11756 |
| b-BISβ | 1.19057 | b-v2 | 0.496814 | 0.630589 | 0.731233 | 0.810825 | b-v3 | 0.413989 |
| γ-BISβ | 1.12972 | γ-v2 | 1.44008 | 2.95958 | 2.69321 | 3.18566 | γ-v3 | 1.81939 |
| φ-BISβ | 1.96824 | φ-v2 | 1.96096 | 2.07987 | 1.96959 | 2.52297 | φ-v3 | 1.06845 |
| ω-BISβ | 4.56578 | ω-v2 | 6.34245 | 6.5076 | 6.80682 | 7.76368 | ω-v3 | 7.07739 |
| b-v1 | 1.0924 | b-v1 | 0.345351 | 0.187181 | 0.140274 | 0.509136 | b-v2 | 0.579945 |
| γ-v1 | 0.80777 | γ-v1 | 2.30501 | 3.71421 | 2.7988 | 3.68344 | γ-v2 | 2.11654 |
| φ-v1 | 0.473447 | φ-v1 | 1.46007 | 1.25789 | 0.695295 | 0.820463 | φ-v2 | 1.84943 |
| ω-v1 | 5.46832 | ω-v1 | 7.41594 | 8.3873 | 8.87475 | 9.44448 | ω-v2 | 8.4093 |
| b-e2 | 0.797867 | b-e2 | 0.243227 | 0.287767 | 0.294475 | 0.305633 | b-e2 | 0.080687 |
| γ-e2 | 0.699407 | γ-e2 | 0.436812 | 0.468487 | 0.499999 | 0.499976 | γ-e2 | 0.726685 |
| φ-e2 | 3.70491 | φ-e2 | 1.236 | 1.40072 | 1.19254 | 0.873475 | φ-e2 | 1.41368 |
| ω-e2 | 5.52906 | ω-e2 | 5.52197 | 5.5352 | 5.45773 | 5.40738 | ω-e2 | 5.59218 |
| b-BSβ | 0.304859 | b-e1 | 0.333759 | 0.265795 | 0.259914 | 0.285452 | b-BISβ | 0.057197 |
| γ-BSβ | 0.673268 | γ-e1 | 0.615 | 0.491162 | 0.499595 | 0.494742 | γ-BISβ | 0.556848 |
| φ-BSβ | 3.14663 | φ-e1 | 2.60082 | 3.63363 | 3.21614 | 3.14106 | φ-BISβ | 0.372285 |
| ω-BSβ | 3.90035 | ω-e1 | 5.18663 | 5.31168 | 5.20609 | 5.19921 | ω-BISβ | 4.41921 |
| Q-v1 | 6.7 | Q-v1 | 3.22 | 2.25 | 3.17 | 2.56 | Q-v2 | 3.98 |
| | | Q-v2 | 4.40 | 2.19 | 2.52 | 2.44 | Q-v3 | 3.88 |
| n$^*$-v1 | 55 | n$^*$-v1 | 126 | 167 | 211 | 283 | n$^*$-v2 | 445 |
| | | n$^*$-v2 | 147 | 215 | 276 | 345 | n$^*$-v3 | 531 |

**Table S2.** Fitting parameters for the five-oscillator model for the 5L-50nm sample on sapphire in the main text. $a_r$ is a background parameter, ω, b, γ, and φ represent the frequency, strength, linewidth, and phase of the oscillators, respectively. α, BISα, BISβ, and BSβ represent the α phonon of $Bi_2Se_3$, the α phonon of BIS, the β phonon of BIS, and the β phonon of $Bi_2Se_3$, respectively. e1 and e2 represent the ENZ1 and ENZ2 modes of BIS, respectively. v1 and v2 represent the VPP1 and VPP2 modes of the HMM, respectively.

|        | 3um     | 0.66um  |        | 1um     |        | 0.5um   |        | 0.25um  |        | 0.35um  |
|--------|---------|---------|--------|---------|--------|---------|--------|---------|--------|---------|
| a      | 0.03    | 0.18552 | ar     | 0.44346 | ar     | 0.39586 | ar     | 0.59785 | ar     | 0.36704 |
| ω-α    | 1.95957 | 1.99387 | ω-α    | 1.97516 | ω-α    | 1.98131 | ω-α    | 1.99816 | ω-α    | 2.01222 |
| b-α    | 0.28974 | 0.35547 | b-α    | 0.47503 | b-α    | 0.075   | b-α    | 1.4609  | b-α    | 1.0929  |
| γ-α    | 0.45819 | 0.79588 | γ-α    | 0.85857 | γ-α    | 0.85464 | γ-α    | 0.20682 | γ-α    | 0.23562 |
| φ-α    | 2.76846 | 2.25161 | φ-α    | 1.96626 | φ-α    | 2.59956 | φ-α    | 1.90624 | φ-α    | 2.45136 |
| b-e1   | 0.31766 | 0.14702 | b-BISβ | 0.25494 | b-BISβ | 0.13536 | b-e1   | 0.09963 | b-e1   | 0.20098 |
| γ-e1   | 0.68007 | 0.43256 | γ-BISβ | 0.71672 | γ-BISβ | 0.6495  | γ-e1   | 0.45515 | γ-e1   | 0.54507 |
| φ-e1   | 1.28867 | 0.37737 | φ-BISβ | 1.86329 | φ-BISβ | 2.09087 | φ-e1   | 1.23138 | φ-e1   | 1.50583 |
| ω-e1   | 4.74413 | 4.9608  | ω-BISβ | 4.22086 | ω-BISβ | 4.10145 | ω-e1   | 4.83473 | ω-e1   | 4.78514 |
| b-e2   | 0.31658 | 0.83061 | b-e2   | 1.32181 | b-e2   | 1.14021 | b-e2   | 1.19204 | b-e2   | 1.14943 |
| γ-e2   | 0.8054  | 0.34814 | γ-e2   | 0.27337 | γ-e2   | 0.25503 | γ-e2   | 0.58288 | γ-e2   | 0.51454 |
| φ-e2   | 0.97119 | 2.32263 | φ-e2   | 1.77126 | φ-e2   | 1.76364 | φ-e2   | 1.36487 | φ-e2   | 2.14217 |
| ω-e2   | 5.29969 | 5.74075 | ω-e2   | 5.75116 | ω-e2   | 5.73294 | ω-e2   | 5.53556 | ω-e2   | 5.43054 |
| b-v1   | 0.25765 | 0.65549 | b-v2   | 1.24651 | b-BSβ  | 0.17049 | b-BISα | 0.04201 |        |         |
| γ-v1   | 1.43565 | 1.59204 | γ-v2   | 0.33003 | γ-BSβ  | 0.8943  | γ-BISα | 0.42552 |        |         |
| φ-v1   | 1.28285 | 0.72322 | φ-v2   | 0.64331 | φ-BSβ  | 3.89036 | φ-BISα | 4.61584 |        |         |
| ω-v1   | 6.7866  | 8.51659 | ω-v2   | 6.58134 | ω-BSβ  | 3.68352 | ω-BISα | 2.67179 |        |         |
| b-BSβ  | 0.16729 | 0.04194 | b-BSβ  | 0.23897 |        |         |        |         |        |         |
| γ-BSβ  | 0.6485  | 0.66091 | γ-BSβ  | 0.88327 |        |         |        |         |        |         |
| φ-BSβ  | 0.9052  | 1.69893 | φ-BSβ  | 3.1896  |        |         |        |         |        |         |
| ω-BSβ  | 3.72582 | 3.87308 | ω-BSβ  | 3.66316 |        |         |        |         |        |         |

**Reference:**

[1] S. C. Lee, S. S. Ng, H. Abu Hassan, Z. Hassan, and T. Dumelow, Opt. Mater. (Amst). **37**, 773 (2014).

[2]     J. S. Blakemore, J. Appl. Phys. **53**, (1982).